\documentclass[10pt]{article}
\usepackage{amsmath,amssymb,tikz}
\usepackage{graphicx}
 \allowdisplaybreaks
\long\def\symbolfootnote[#1]#2{\begingroup%
\def\thefootnote{\fnsymbol{footnote}}\footnote[#1]{#2}\endgroup}

\newcommand{\aei}{\it Max Planck Institute for Gravitational Physics
(Albert Einstein Institute)\\ Am M\"uhlenberg 1, 14476 Golm,
Germany}

\newcommand{\ictsustc}{\it Interdisciplinary Center for Theoretical Study,
University of Science and Technology of China,\\
Hefei, Anhui 230026, People's Republic of China.}
\begin{document}
\thispagestyle{empty}
\begin{flushright}
\hfill{AEI-2014-003}
\end{flushright}
\begin{center}

~\vspace{20pt}

{\Large\bf A Note on Holographic Weyl Anomaly and Entanglement
Entropy}

\vspace{25pt}

Rong-Xin Miao\symbolfootnote[1]{Email:~\sf rong-xin.miao@aei.mpg.de}

\vspace{10pt}${}^\ast{}$\aei

\vspace{10pt}${}^\ast{}$\ictsustc

\vspace{2cm}

\begin{abstract}
We develop a general approach to simplify the derivation of the
holographic Weyl anomaly. As an application, we derive the
holographic Weyl anomaly from general higher derivative gravity in
asymptotically $AdS_{5}$ and $AdS_{7}$. Interestingly, to derive all
the central charges of 4d and 6d CFTs, we make no use of equations
of motion. Following Myers' idea, we propose a formula of
holographic entanglement entropy for higher derivative gravity in
asymptotically $AdS_5$. Applying this formula, we obtain the correct
universal term of entanglement entropy for 4d CFTs. It turns out that our formula is the leading term of Dong's proposal in asymptotically $AdS_5$. Since only the leading term contributes to the universal log term, we actually prove that Dong's proposal yields the correct universal term of entanglement entropy for 4d CFTs. This is a nontrivial test of Dong's proposal.
\end{abstract}

\end{center}

 \newpage

\tableofcontents

\section{Introduction}
The AdS/CFT correspondence \cite{Maldacena} is an exact realization
of the holographic principle \cite{Hooft,Susskind,Bousso}, which
claims that the quantum gravity theory in the bulk is dual to the
gauge field theory on the boundary. It provides a powerful tool to
study the nonperturbative phenomena of gauge theories
\cite{Maldacena1}.

An interesting test of AdS/CFT correspondence is the successful
derivation of the holographic Weyl anomaly from gravity theories. It
was firstly proposed by Witten \cite{Witten} and then worked out in
detail by Henningson et al for Einstein gravity \cite{Henningson}.
Applying the so-called ``PBH transformation'' (relation between
diffeomorphisms in the bulk and Weyl transformation on the
boundary), Imbimbo et al obtain a universal formula for the type A
anomaly ( which is related to the Euler characteristic) for higher
derivative gravity \cite{Theisen}. Interestingly, they make no use
of equations of motion. While for the type B anomaly, there is no
universal formula for higher derivative gravity so far. For
interesting developments of the holographic Weyl anomaly, please
refer to
\cite{Nojiri,Nojiri1,Nojiri2,Nojiri3,Mir,Bastianelli,deBoer, Fukuma,
Kraus}. For a good review of the Weyl anomaly, please refer to
\cite{Duff}. See also \cite{Boulanger,Boulanger1,Boulanger2} for the
general structure of the Weyl anomaly.

In this note, we try to develop a simple approach to derive the
holographic Weyl anomaly from general higher derivative gravity. We
firstly expand the action around a referenced curvature, then select
and calculate the terms relevant to the Weyl anomaly. Interestingly,
we only need to calculate very few terms after expanding the action,
which highly simplifies calculations. Remarkably, there are only two
(four) relevant terms in five (seven) dimensional spacetime, which
is just the number of independent central charges of the
corresponding CFTs. Applying our approach, we derive the general
formulas of type B anomaly from higher derivative gravity in
asymptotically $AdS_{5}$ and $AdS_{7}$. Interestingly, we make no
use of equations of motion to obtain all the charges of 4d and 6d
CFTs. However, it is expected that one has to solve equations of
motion for the type B anomaly in higher dimensions.

As an application of our general formulas, we propose a formula of
holographic entanglement entropy \cite{Ryu1,Ryu2,Ryu3} for higher
derivative gravity in asymptotically $AdS_5$. We prove that it
yields the correct logarithmic term of the entanglement entropy for
4d CFTs. Besides, it is consistent with the formula of holographic
entanglement entropy for Love-Lock gravity \cite{Boer,Hung}, the
curvature-squared gravity \cite{Solodukhin1} and recent proposals of Dong \cite{Dong} and Camps \cite{Camps}. We find that our formula is the leading term of Dong's proposal in asymptotically $AdS_{5}$. Since only the leading term contributes to the universal log term, we actually prove that Dong's proposal yields the correct universal term of entanglement entropy for 4d CFTs. This is a nontrivial test of Dong's proposal. For other recent developments of the holographic entanglement entropy, please refer to \cite{Lewkowycz1,Lewkowycz2,Chen,Bhattacharyya1,Bhattacharyya2}.

The paper is organized as follows. In Sect. 2, we develop a general
approach to simplify the calculations of the holographic Weyl
anomaly from higher derivative gravity. We derive the universal
formulas of the holographic Weyl anomaly for 4d and 6d CFTs. In
Sect. 3, we study some examples to show the application of our
general approach. In Sect. 4, we propose a formula of holographic
entanglement entropy in asymptotically $AdS_5$. We conclude in Sect.
5.

\section{Holographic Weyl Anomaly}

In this section, we develop a simple approach to derive the
holographic Weyl anomaly from general higher derivative gravity in
$AdS/CFT$ correspondence. The main idea is as follows. Firstly we
expand the action around a referenced curvature, then select and
calculate the terms relevant to the holographic Weyl anomaly. For simplicity, we list the complicated formulas of Riemann tensors and the referenced curvature in the appendix.  We find they are useful in our following discussions.

Let us consider the higher derivative gravity with the action
\begin{eqnarray}\label{action}
S=\frac{1}{2\kappa^2_{d+1}}\int
d^{d+1}x\sqrt{-\hat{G}}f(\hat{R}_{\mu\nu\rho\sigma})+S_B,
\end{eqnarray}
where $f(\hat{R}_{\mu\nu\rho\sigma})$ is a scalar function
constructed from the curvature, $S_B$ is the boundary term for a
well-defined variational principle. For simplicity, we focus on the
case that $f(\hat{R}_{\mu\nu\rho\sigma})$ contains no derivatives of
the curvatures. Our discussions can be easily generalized to the
case with derivatives of curvatures. We study those cases in some
examples. We also ignore $S_B$ in the following discussions since it
does not contribute to the Weyl anomaly. From eq.(\ref{action}), we
can derive the equations of motion as
\begin{eqnarray}\label{fieldequation}
P_{\mu}^{\
\alpha\rho\sigma}\hat{R}_{\nu\alpha\rho\sigma}-2\nabla^{\rho}\nabla^{\sigma}P_{\mu\rho\sigma\nu}-\frac{1}{2}f\hat{G}_{\mu\nu}=0,
\end{eqnarray}
with $P^{\mu\nu\rho\sigma}=\delta f/\delta
\hat{R}_{\mu\nu\rho\sigma}$. We assume eq.(\ref{fieldequation}) has
an asymptotically $AdS$ solution with the metric
\begin{eqnarray}\label{metric}
ds^2=\hat{G}_{\mu\nu}dx^{\mu}dx^{\nu}=\frac{1}{4\rho^2}d\rho^2+\frac{1}{\rho}g_{ij}dx^idx^j,
\end{eqnarray}
where $g_{ij}=g_{(0)ij}+\rho g_{(1)ij}+...+\rho^{\frac{d}{2}}(
g_{(\frac{d}{2})ij}+h_{\frac{d}{2}}\log \rho)+...$ when $d$ is even. 

Now let us begin to derive the holographic Weyl anomaly. Using the
asymptotically AdS solution eq.(\ref{metric}), we can expand the
action as
\begin{eqnarray}\label{action1}
2\kappa^2_{d+1}S&=&\int
d^{d+1}x\sqrt{-\hat{G}}f(\hat{R}_{\mu\nu\rho\sigma})=\frac{1}{2}\int
d\rho d^dx\rho^{-\frac{d}{2}-1}\sqrt{-g_{(0)}}b(x,\rho),\nonumber\\
&&b(x,\rho)=b_{0}(x)+\rho b_{1}(x)+\rho^2 b_2 (x)+...
\end{eqnarray}
According to \cite{Theisen}, the holographic Weyl anomaly is
\begin{eqnarray}
<T^i_i>=\frac{1}{2\kappa^2_{d+1}}b_{\frac{d}{2}},
\end{eqnarray}
with d an even number. By dimensional analysis, we note that
$b_{2m}$ contains the square of $g_{(m)ij}$. So we can derive
equations of motion of $g_{(m)ij}$ from the variation of
$\sqrt{-g_{(0)}}b_{2m}$ ($m>0$). Besides, $b_{m+1}$ contains only
linear terms of $g_{([\frac{m+1}{2}]+1)ij},...,g_{(m+1)ij}$. Using
equations of motion, all these linear terms vanish.

Let us expand $f$ around a referenced curvature
$\bar{R}_{\mu\nu\rho\sigma}=-(\hat{G}_{\mu\rho}\hat{G}_{\nu\sigma}-\hat{G}_{\mu\sigma}\hat{G}_{\nu\rho})$
:
\begin{eqnarray}\label{f}
f(\hat{R}_{\mu\nu\rho\sigma})&=&f(\bar{R})+P^{\mu\nu\rho\sigma}|_{\bar{R}}(\hat{R}-\bar{R})_{\mu\nu\rho\sigma}\nonumber\\
&+&\frac{1}{2}\frac{\delta^2f}{\delta\hat{R}_{\mu\nu\rho\sigma}\delta\hat{R}_{\mu_1\nu_1\rho_1\sigma_1}}|_{\bar{R}}(\hat{R}-\bar{R})_{\mu\nu\rho\sigma}(\hat{R}-\bar{R})_{\mu_1\nu_1\rho_1\sigma_1}\nonumber\\
&+&\frac{1}{3!}\frac{\delta^3f}{\delta\hat{R}_{\mu\nu\rho\sigma}\delta\hat{R}_{\mu_1\nu_1\rho_1\sigma_1}\delta\hat{R}_{\mu_2\nu_2\rho_2\sigma_2}}|_{\bar{R}}(\hat{R}-\bar{R})_{\mu\nu\rho\sigma}(\hat{R}-\bar{R})_{\mu_1\nu_1\rho_1\sigma_1}(\hat{R}-\bar{R})_{\mu_2\nu_2\rho_2\sigma_2}\nonumber\\
&+&...
\end{eqnarray}
Notice that the referenced curvature is different from the
asymptotically AdS curvature
$-(\hat{G}_{(0)\mu\rho}\hat{G}_{(0)\nu\sigma}-\hat{G}_{(0)\mu\sigma}\hat{G}_{(0)\nu\rho})$
with $\hat{G}_{(0)00}=\frac{1}{4\rho2},\
\hat{G}_{(0)ij}=\frac{1}{\rho}g_{(0)ij}$. For useful properties of the referenced curvature, please refer to eqs.(\ref{referenceR}-\ref{order3}) in the appendix. Let us denote the $n$-th
order of Taylor expansions by $f_n$
\begin{eqnarray}\label{fn}
f_n=\frac{1}{n!}\frac{\delta^nf}{\delta\hat{R}_{\mu_1\nu_1\rho_1\sigma_1}...\delta\hat{R}_{\mu_n\nu_n\rho_n\sigma_n}}|_{\bar{R}}(\hat{R}-\bar{R})_{\mu_1\nu_1\rho_1\sigma_1}...(\hat{R}-\bar{R})_{\mu_n\nu_n\rho_n\sigma_n}.
\end{eqnarray}
According to eqs.(\ref{order1}-\ref{order3}) in the appendix, we find that $f_n$ behaves at least
as order $o(\rho^n)$. So to derive the holographic Weyl anomaly in d
dimensions, we only need to consider the terms up to the
$\frac{d}{2}$-th order $(f_0, f_1,..., f_{\frac{d}{2}})$. In
general, we have
\begin{eqnarray}\label{Expandvector}
\frac{1}{n!}\frac{\delta^{n}f}{\delta
\hat{R}^{\mu_1\nu_1\rho_1\sigma_1}...\delta
\hat{R}^{\mu_{n}\nu_{n}\rho_{n}\sigma_{n}}}|_{\bar{R}}=\sum_{i=1}^{m_n}c^n_i
X^{n}_{\ i \
\mu_1\nu_1\rho_1\sigma_1,...,\mu_{n}\nu_{n}\rho_{n}\sigma_{n}},
\end{eqnarray}
where $c^n_i$ are constants and $m_n$ is the number of independent
scalars constructed from appropriate contractions of $n$ curvature tensors. For example, $m_1=1, m_2=3, m_3=8$. Tensor $X^n_i$
is defined as
\begin{eqnarray}\label{TensorX}
X^n_{\ i \
\mu_1\nu_1\rho_1\sigma_1,...,\mu_{n}\nu_{n}\rho_{n}\sigma_{n}}=\frac{1}{n!}\frac{\delta^{n}K^n_i}{\delta
\hat{R}^{\mu_1\nu_1\rho_1\sigma_1}...\delta
\hat{R}^{\mu_{n}\nu_{n}\rho_{n}\sigma_{n}}},
\end{eqnarray}
with $K^n_i$ denotes the independent scalars constructed from $n$
curvature tensors. For example, we have
\begin{eqnarray}\label{Kni}
&&K^1_1=\hat{R},\nonumber\\
&&K^2_i=(\hat{R}_{\mu\nu\rho\sigma}\hat{R}^{\mu\nu\rho\sigma},\
\hat{R}_{\mu\nu}\hat{R}^{\mu\nu},\
\hat{R}^2), \nonumber\\
&&K^3_i=(\hat{R}^3,\hat{R}\hat{R}_{\mu\nu}\hat{R}^{\mu\nu},\hat{R}\hat{R}_{\mu\nu\rho\sigma}\hat{R}^{\mu\nu\rho\sigma},\hat{R}_{\mu}^{\nu}\hat{R}_{\nu}^{\rho}\hat{R}_{\rho}^{\mu},
\hat{R}^{\mu\nu}\hat{R}^{\rho \sigma}\hat{R}_{\mu \rho \sigma\nu},
\hat{R}_{\mu \nu}\hat{R}^{\mu \rho \sigma \lambda}\hat{R}^{\nu}_{\ \rho \sigma \lambda},\nonumber\\
&&\ \ \ \ \ \ \ \ \hat{R}_{\mu \nu \rho \sigma}\hat{R}^{\mu \nu
\lambda\chi}\hat{R}^{\rho \sigma}_{\ \ \lambda\chi},\hat{R}_{\nu \nu
\rho\sigma}\hat{R}^{\nu\lambda\chi\sigma}\hat{R}^{\nu\ \ \rho}_{\
\lambda\chi}),\nonumber
\\&&...
\end{eqnarray}
Applying eqs.(\ref{Expandvector},\ref{TensorX},\ref{Kni}), we can
rewrite $f_n$ in a very nice form as
\begin{eqnarray}\label{generalfn}
f_n=\sum_{i=1}^{m_n}c^n_i \tilde{K}^n_i,
\end{eqnarray}
with $\tilde{K}^n_i=K^n_i|_{[\hat{R}\rightarrow(\hat{R}-\bar{R})]}$
and $c^n_i $ is determined by the action. It should be mentioned
that not all of $\tilde{K}^n_i (n\leq \frac{d}{2})$ contribute to
the holographic Weyl anomaly. Applying
eqs.(\ref{order1},\ref{order2}), we can select the terms relevant to
the Weyl anomaly.

Using the assumption that the higher derivative gravity has an
asymptotically $AdS$ solution, we can prove $c^1_1=-\frac{f_0}{2d}$.
We show the proof below. From the above equations, we can derive
\begin{eqnarray}\label{P}
P^{\mu\nu\rho\sigma}|_{\bar{R}}= \frac{c^1_1}{2}
(\hat{G}^{\mu\rho}\hat{G}^{\nu\sigma}-\hat{G}^{\mu\sigma}\hat{G}^{\nu\rho}).
\end{eqnarray}
 A useful formula in the
following derivations is
\begin{eqnarray}\label{g1}
g_{(1)ij}=-\frac{1}{d-2}(R_{(0)ij}-\frac{R_{0}}{2(d-1)}g_{(0)ij}),
\end{eqnarray}
which is determined completely by PBH transformation \cite{Theisen}
and independent of equations of motion. Based on our above
discussions, we can derive $g_{(1)ij}$ from the variation of
$b_{2}$. To get $b_{2}$, we only need to consider terms $f_0, f_1,
f_2$. As we shall show in the next section, eq.(\ref{g1}) can be
derived from $\sqrt{-\hat{G}}f_2$ independently. Thus, we must be
able to derive eq.(\ref{g1}) from $\sqrt{-\hat{G}}(f_0+f_1)$. Using
eq.(\ref{P}), we have
\begin{eqnarray}\label{f0f1}
\sqrt{-\hat{G}}(f_0+f_1)=c^1_1\sqrt{-\hat{G}}(\hat{R}+d^2+d+\frac{f_0}{c^1_1}).
\end{eqnarray}
Compared with the Einstein-Hilbert action with a negative
cosmological constant
\begin{eqnarray}\label{EH}
\sqrt{-\hat{G}}(\hat{R}-2\Lambda)=\sqrt{-\hat{G}}(\hat{R}+d^2-d),
\end{eqnarray}
it is clear that $c^1_1=-\frac{f_0}{2d}$ is the only solution which
can yield the correct expression of $g_{(1)ij}$ eq.(\ref{g1}). Thus,
in general, we have
\begin{eqnarray}\label{cP}
P^{\mu\nu\rho\sigma}|_{\bar{R}}= -\frac{f_0}{4d}
(\hat{G}^{\mu\rho}\hat{G}^{\nu\sigma}-\hat{G}^{\mu\sigma}\hat{G}^{\nu\rho}),\\
\sqrt{-\hat{G}}(f_0+f_1)=-\frac{f_0}{2d}\sqrt{-\hat{G}}(\hat{R}+d^2-d).\label{cf0f1}
\end{eqnarray}
Actually, there is a simple method to derive $c^1_1$. Suppose that
AdS is an exact solution to eq.(\ref{fieldequation}), then the
curvature $\hat{R}_{\mu\nu\rho\sigma}$ becomes exactly the
referenced curvature $\bar{R}_{\mu\nu\rho\sigma}$. Substituting
$\bar{R}_{\mu\nu\rho\sigma}$ and eq.(\ref{P}) into
eq.(\ref{fieldequation}), we can derive $c^1_1=-\frac{f_0}{2d}$
directly.

To summarize, we list the main steps of our approach. Firstly, we
expand the action around the referenced curvature (\ref{referenceR})
up to the $\frac{d}{2}$-order
\begin{equation}
f=\sum_{n=0}^{\frac{d}{2}}
f_n=\sum_{n=0}^{\frac{d}{2}}\sum_{i=1}^{m_n} c^n_i\tilde{K}^n_i.
\end{equation}
Secondly, we select the terms relevant to the Weyl anomaly with the
help of eqs.(\ref{order1}, \ref{order2}). As we shall show in the
following sections, only very few terms contribute to the Weyl
anomaly. Finally, we calculate these relevant terms to derive the
holographic Weyl anomaly.

\subsection{4d Weyl Anomaly}
In this subsection, we derive the holographic Weyl anomaly for 4d
CFTs. As discussed in the above section, we only need to consider
the terms $\sqrt{-\hat{G}}(f_0+f_1+f_2)$ for the calculations of the
Weyl anomaly $<T^i_i>=\frac{1}{2\kappa^2_{5}}b_2$.

Applying eq.(27) of \cite{Nojiri} and eqs.(\ref{g1},\ref{cf0f1}), we
can derive
\begin{eqnarray}\label{A}
\sqrt{\hat{G}}(f_0+f_1)=\frac{\sqrt{g_{(0)}}}{2\rho}\frac{f_0}{64}(E_4-C_{ijkl}C^{ijkl})+...
\end{eqnarray}
$``..."$ in this paper denotes the total derivative or terms
irrelevant to the Weyl anomaly. $E_4=R_{(0)ijkl}R_{(0)}^{\ \
ijkl}-4R_{(0)ij}R_{(0)}^{\ \ ij}+R_{(0)}^{\ \ 2}$ and
$C_{ijkl}C^{ijkl}=R_{(0)ijkl}R_{(0)}^{\ \ ijkl}-2R_{(0)ij}R_{(0)}^{\
\ ij}+\frac{1}{3}R_{(0)}^{\ \ 2}$ are the four-dimensional Euler
density and square of Weyl tensor, respectively.

Now let us go on to compute $\sqrt{-\hat{G}}f_2$. From
eqs.(\ref{Expandvector}, \ref{TensorX},\ref{Kni}), we have
\begin{eqnarray}\label{C2} \frac{1}{2}\frac{\delta^2f}{\delta
\hat{R}_{\mu\nu\rho\sigma}\delta
\hat{R}^{\mu_1\nu_1\rho_1\sigma_1}}|_{\bar{R}}=c^2_1X^{2\ \
\mu\nu\rho\sigma}_{1 \mu_1\nu_1\rho_1\sigma_1}+c^2_2X^{2\ \
\mu\nu\rho\sigma}_{2 \mu_1\nu_1\rho_1\sigma_1}+c^2_3X^{2\ \
\mu\nu\rho\sigma}_{3 \mu_1\nu_1\rho_1\sigma_1},
\end{eqnarray}
where $X^2_1,X^2_2,X^2_3$ are three independent tensors defined as
follows:
\begin{eqnarray}\label{X1}
X^{2\ \mu\nu\rho\sigma}_{1 \mu_1\nu_1\rho_1\sigma_1}=\frac{\ \
\partial \hat{R}_{\mu_1\nu_1\rho_1\sigma_1}}{\partial
\hat{R}_{\mu\nu\rho\sigma}}=&&\frac{1}{12}(\delta^{\mu\nu}_{\mu_1\nu_1}\delta^{\rho\sigma}_{\rho_1\sigma_1}-\frac{1}{2}\delta^{\mu\rho}_{\mu_1\nu_1}\delta^{\sigma\nu}_{\rho_1\sigma_1}-\frac{1}{2}\delta^{\mu\sigma}_{\mu_1\nu_1}\delta^{\nu\rho}_{\rho_1\sigma_1}\nonumber\\
&+&\delta^{\mu\nu}_{\rho_1\sigma_1}\delta^{\rho\sigma}_{\mu_1\nu_1}-\frac{1}{2}\delta^{\mu\rho}_{\rho_1\sigma_1}\delta^{\sigma\nu}_{\mu_1\nu_1}-\frac{1}{2}\delta^{\mu\sigma}_{\rho_1\sigma_1}\delta^{\nu\rho}_{\mu_1\nu_1})
\end{eqnarray}
\begin{eqnarray}\label{X2}
X^{2\ \mu\nu\rho\sigma}_{2
\mu_1\nu_1\rho_1\sigma_1}=\frac{1}{4}\hat{G}_{\alpha\beta}(X^{2\
\mu\alpha\rho\beta}_{1\mu_1\nu_1\rho_1\sigma_1}\hat{G}^{\nu\sigma}-X^{2\
\mu\alpha\sigma\beta}_{1\mu_1\nu_1\rho_1\sigma_1}\hat{G}^{\nu\rho}-X^{2\
\nu\alpha\rho\beta}_{1\mu_1\nu_1\rho_1\sigma_1}\hat{G}^{\mu\sigma}+X^{2\
\nu\alpha\sigma\beta}_{1\mu_1\nu_1\rho_1\sigma_1}\hat{G}^{\mu\rho})\nonumber\\
\end{eqnarray}
\begin{eqnarray}\label{X3}
X^{2\ \mu\nu\rho\sigma}_{3
\mu_1\nu_1\rho_1\sigma_1}=\frac{1}{4}(\hat{G}^{\mu\rho}\hat{G}^{\nu\sigma}-\hat{G}^{\mu\sigma}\hat{G}^{\nu\rho})(\hat{G}_{\mu_1\rho_1}\hat{G}_{\nu_1\sigma_1}-\hat{G}_{\mu_1\sigma_1}\hat{G}_{\nu_1\rho_1})
\end{eqnarray}
Here we have
$\delta^{\mu\nu}_{\mu_1\nu_1}=\delta^{\mu}_{\mu_1}\delta^{\nu}_{\nu_1}-\delta^{\mu}_{\nu_1}\delta^{\nu}_{\mu_1}$.
 Let us define a new tensor $Y$ for $d=4$
\begin{eqnarray}\label{Y}
Y=\frac{1}{210}(6X^2_1-8X^2_2+X^2_3),\ \ Y*X^2_1=1,\
Y*X^2_2=Y*X^2_3=0,
\end{eqnarray}
where $Y*X=Y^{\ \mu\nu\rho\sigma}_{\mu_1\nu_1\rho_1\sigma_1}X_{ \
\mu\nu\rho\sigma}^{\mu_1\nu_1\rho_1\sigma_1}$.

Using eq.(27) of \cite{Nojiri} and eq.(\ref{g1}), we can derive
\begin{eqnarray}\label{XX1}
&&\sqrt{\hat{G}}X^{2\ \mu\nu\rho\sigma}_{1
\mu_1\nu_1\rho_1\sigma_1}(\hat{R}_{\mu\nu\rho\sigma}-\bar{R}_{\mu\nu\rho\sigma})(\hat{R}_{\mu_1\nu_1\rho_1\sigma_1}-\bar{R}_{\mu_1\nu_1\rho_1\sigma_1})\nonumber\\
&&=\sqrt{\hat{G}}(\hat{R}_{\mu\nu\rho\sigma}-\bar{R}_{\mu\nu\rho\sigma})(\hat{R}^{\mu\nu\rho\sigma}-\bar{R}^{\mu\nu\rho\sigma})\nonumber\\
&&=\sqrt{\hat{G}}(\hat{R}_{\mu\nu\rho\sigma}\hat{R}^{\mu\nu\rho\sigma}+4\hat{R}+2d(d+1))\nonumber\\
&&=\frac{\sqrt{g_{(0)}}}{2\rho}C_{ijkl}C^{ijkl}+o(1).
\end{eqnarray}
A useful formula in the above derivation is $X^{\
\mu\nu\rho\sigma}_{1
\mu_1\nu_1\rho_1\sigma_1}Z^{\mu_1\nu_1\rho_1\sigma_1}=Z^{\mu\nu\rho\sigma}$
where $Z^{\mu\nu\rho\sigma}$ has the same symmetry properties as
$\hat{R}^{\mu\nu\rho\sigma}$. Following the same methods, one can
derive
\begin{eqnarray}\label{XX2}
&&\sqrt{\hat{G}}X^{2\ \mu\nu\rho\sigma}_{2
\mu_1\nu_1\rho_1\sigma_1}(\hat{R}_{\mu\nu\rho\sigma}-\bar{R}_{\mu\nu\rho\sigma})(\hat{R}_{\mu_1\nu_1\rho_1\sigma_1}-\bar{R}_{\mu_1\nu_1\rho_1\sigma_1})\nonumber\\
&&=\sqrt{\hat{G}}(\hat{R}_{\mu\nu}-\bar{R}_{\mu\nu})(\hat{R}^{\mu\nu}-\bar{R}^{\mu\nu})\nonumber\\
&&=o(\rho),
\end{eqnarray}
\begin{eqnarray}\label{XX3}
&&\sqrt{\hat{G}}X^{2\ \mu\nu\rho\sigma}_{3
\mu_1\nu_1\rho_1\sigma_1}(\hat{R}_{\mu\nu\rho\sigma}-\bar{R}_{\mu\nu\rho\sigma})(\hat{R}_{\mu_1\nu_1\rho_1\sigma_1}-\bar{R}_{\mu_1\nu_1\rho_1\sigma_1})\nonumber\\
&&=\sqrt{\hat{G}}(\hat{R}-\bar{R})^2\nonumber\\
&&=o(\rho),
\end{eqnarray}
which do not contribute to the holographic Weyl anomaly. In the
above calculations, we have used eq.(\ref{order2}). As mentioned in
the above section, we can derive eq.(\ref{g1}) form the variation of
the third line of eq.(\ref{XX1}) with respect to $g_{(1)ij}$.

Now we obtain
\begin{eqnarray}\label{f2}
\sqrt{\hat{G}}f_2=\frac{\sqrt{g_{(0)}}}{2\rho}c^2_1C_{ijkl}C^{ijkl}+o(1).
\end{eqnarray}
From eqs.(\ref{C2},\ref{Y}), we have
\begin{eqnarray}\label{n1}
c^2_1=\frac{1}{2}\frac{\delta^2f}{\delta
\hat{R}_{\mu\nu\rho\sigma}\delta
\hat{R}^{\mu_1\nu_1\rho_1\sigma_1}}|_{\bar{R}}\  Y_{\
\mu\nu\rho\sigma}^{ \mu_1\nu_1\rho_1\sigma_1}.
\end{eqnarray}
Combining eqs.(\ref{A},\ref{f2}), we get
\begin{eqnarray}\label{b2}
b_2=\frac{f_0}{64}E_4-(\frac{f_0}{64}-c^2_1)C_{ijkl}C^{ijkl}.
\end{eqnarray}
So the holographic Weyl anomaly for 4d CFT is
\begin{eqnarray}\label{4dWeyl}
<T^i_i>=\frac{1}{2\kappa^2_{d+1}}b_2=\frac{c}{16\pi^2}C_{ijkl}C^{ijkl}-\frac{a}{16\pi^2}E_4,
\end{eqnarray}
with
\begin{eqnarray}\label{ac}
a=-\frac{f_0}{8}\frac{\pi^2}{\kappa_5^2}, \ \ \
c=(8c^2_1-\frac{f_0}{8})\frac{\pi^2}{\kappa_5^2}.
\end{eqnarray}
Note that we have $\hat{R}=-d(d+1)$ and thus $f_0<0$. As an simple
example, one can check that our formula eq.(\ref{ac}) yields the
correct central charges (eq.(6.5) of \cite{Hung}) for the
curvature-squared action. Our formula is more general, it can apply
to any higher derivative gravity with an asymptotically AdS
solution.

Note that $c^2_1$ is the number of $R_{ijkl}R^{ijkl}$ included in
$b_2$. From eqs.(\ref{iii},\ref{iv},\ref{v}), we observe that it is
$\hat{R}_{\mu\nu\rho\sigma}$ rather than $\hat{R}_{\mu\nu}$ and
$\hat{R}$ that contributes to $R_{ijkl}$. One can also find that
$g_{(n)}$ with $n>1$ do not contribute to $b_2$, while $g_{(1)}$
eq.(\ref{g1}) is independent of $R_{ijkl}$. So
$\hat{R}_{\mu\nu\rho\sigma}$ is the only term that can contribute to
$R_{ijkl}$. Thus $c^2_1$ vanishes if $f(\hat{R}_{\mu\nu\rho\sigma})$
is made of scalars with less than two $\hat{R}_{\mu\nu\rho\sigma}$.
For example, $c^2_1=0$ for
$f(\hat{R},\hat{R}_{\mu\nu}\hat{R}^{\mu\nu},
\hat{R}_{\mu\nu}\hat{R}_{\rho\sigma}\hat{R}^{\mu\nu\rho\sigma})$. In
other words,
$f(\hat{R},\hat{R}_{\mu\nu}\hat{R}^{\mu\nu},\hat{R}_{\mu\nu}\hat{R}_{\rho\sigma}\hat{R}^{\mu\nu\rho\sigma})$
gravity has the same $a$ charge and $c$ charge.

\subsection{6d Weyl Anomaly}

In this section, we derive the holographic Weyl anomaly for 6d CFTs.
We need to calculate $b_3$. Only terms $f_0, f_1, f_2, f_3$ will
contribute to $b_3$. Besides, because $b_3$ only contains terms
linear with $g_{(2)ij}$ which vanish on shell, so we do not need
$g_{(2)ij}$ for the derivations of $b_3$. This means that we do not
need to solve equations of motion in order to derive the holographic
Weyl anomaly for 6d CFT. We have checked straightly that the
$g_{(2)ij}$ terms in $b_3$ indeed vanish after imposing
eq.(\ref{g1}).

Let us list the Weyl invariant quantities in 6 dimensions:
\begin{eqnarray}\label{6dWeylinvariant}
I_1&=&C_{kijl}C^{imnj}C_{m\ \ \ n}^{\ \ kl},\
I_2=C_{ij}^{\ \ kl}C_{kl}^{\ \ mn}C_{mn}^{\ \ \ ij},\nonumber\\
I_3&=&C_{iklm}(\nabla^2\delta^i_j+4R^i_j-\frac{6}{5}R\delta^i_j)C^{jklm}\nonumber\\
E_0&=&384\pi^3
E_6=K^3_1-12K^3_2+3K^3_3+16K^3_4-24K^3_5-24K^3_6+4K^3_7+8K^3_8,
\end{eqnarray}
where $K^3_i$ is defined as
\begin{eqnarray}\label{K3i}
K^3_i=(&&R^3,RR_{ij}R^{ij},RR_{ijkl}R^{ijkl},R_i^jR_j^kR_k^i,
R^{ij}R^{kl}R_{iklj},
R_{ij}R^{iklm}R^j_{klm},\nonumber\\
&&R_{ijkl}R^{ijmn}R^{kl}_{mn},R_{ijkl}R^{imnl}R^{j\ \ \ k}_{mn}).
\end{eqnarray}

Firstly, let us compute the term $\sqrt{-\hat{G}}(f_0+f_1)$:
\begin{eqnarray}\label{6df0f1}
\sqrt{\hat{G}}(f_0+f_1)&=&-\frac{f_0}{2d}\sqrt{-\hat{G}}(\hat{R}+d^2-d)\nonumber\\
&=&-\frac{f_0}{2d}\frac{\sqrt{-\hat{g_{(0)}}}}{2\rho}\frac{1}{192}(E_0-12I_1-3I_2+I_3)+...
\end{eqnarray}

Next, let us calculate $\sqrt{-\hat{G}}f_2$. From
eqs.(\ref{generalfn},\ref{order2}), we note that
\begin{eqnarray}\label{6dX2}
&&\tilde{K}^2_2=(\hat{R}-\bar{R})_{\mu\nu}(\hat{R}-\bar{R})^{\mu\nu}=o(\rho^4),\nonumber\\
&&\tilde{K}^2_3=(\hat{R}-\bar{R})^2=o(\rho^4),
\end{eqnarray}
which do not contribute to $b_3$. So we only need to calculate the
$\tilde{K}^2_1$ term for the holographic Weyl anomaly. After a long
calculation, we obtain

\begin{eqnarray}\label{6df2}
\sqrt{\hat{G}}f_2&=&c^2_1\sqrt{\hat{G}}(\hat{R}-\bar{R})_{\mu\nu\rho\sigma}(\hat{R}-\bar{R})^{\mu\nu\rho\sigma}+...\nonumber\\
&=&c^2_1\frac{\sqrt{-\hat{g_{(0)}}}}{2\rho}(-\frac{1}{3}I_1+\frac{1}{12}I_2+\frac{1}{12}I_3)+...
\end{eqnarray}

Finally, let us calculate the last term $\sqrt{-\hat{G}}f_3$. Using
eqs.(\ref{generalfn},\ref{order2}), we have
\begin{eqnarray}\label{orderf3}
\tilde{K}^3_1&=&o(\rho^6),\ \tilde{K}^3_2=o(\rho^6),\
\tilde{K}^3_3=o(\rho^4),\ \tilde{K}^3_4=o(\rho^6),\
\tilde{K}^3_5=o(\rho^5),\ \tilde{K}^3_6=o(\rho^4),\nonumber\\
\tilde{K}^3_7&=&o(\rho^3),\ \tilde{K}^3_8=o(\rho^3).
\end{eqnarray}
Focus on the $o(\rho^3)$ terms which contribute to Weyl anomaly, we
only need to calculate the terms $\tilde{K}^3_7,\ \tilde{K}^3_8$.
After a complicated calculation, we get
\begin{eqnarray}\label{6dK37}
\sqrt{-\hat{G}}\tilde{K}^3_7=\frac{\sqrt{-g_{(0)}}}{2\rho}I_2,\nonumber\\
\sqrt{-\hat{G}}\tilde{K}^3_8=\frac{\sqrt{-g_{(0)}}}{2\rho}I_1.
\end{eqnarray}

Combining eqs.(\ref{6df0f1},\ref{6df2}), we can derive the
holographic Weyl anomaly for 6d CFT as
\begin{eqnarray}\label{6dWeylanomaly}
<T^i_i>=\frac{1}{2\kappa^2_{7}}b_3=\sum_{n=1}^3B_nI_n+2A E_6,
\end{eqnarray}
with
\begin{eqnarray}\label{6dAB}
&&2\kappa^2_{7}A=-\frac{\pi^3}{12}f_0,\nonumber\\
&&2\kappa^2_{7}B_1=-\frac{1}{3}c^2_1+c^3_8+\frac{1}{192}f_0,\nonumber\\
&&2\kappa^2_{7}B_2=\frac{1}{12}c^2_1+c^3_7+\frac{1}{768}f_0,\nonumber\\
&&2\kappa^2_{7}B_3=\frac{1}{12}c^2_1-\frac{1}{2304}f_0.
\end{eqnarray}
The calculations of the coefficients $c^2_1, c^3_7$ and $c^3_8$ for
the general action are quite complicated. We list the main steps and
results in the Appendix. While for a given action, as we shall show
in the next section, we can always get these coefficients easily by
expanding the action directly.

It is interesting that only four independent coefficients ($f_0,
c^2_1, c^3_7, c^3_8$) contribute to the Weyl anomaly which exactly
agrees with the number of independent central charges of CFT. It is
also remarkable that, similar to the 4d case, only
$\sqrt{-G}(f_0+f_1)$ rather than $\sqrt{-G}(f_2+f_3+...)$ contribute
to the central charge with respect to the Euler density. In fact,
this is a general conclusion. According to \cite{Theisen}, the type
A trace anomaly from general gravity action is
\begin{eqnarray}\label{typeAanmoly}
b_{\frac{d}{2}}=\frac{f_0}{(d!)^2}E_{2n}+...
\end{eqnarray}
where ``...'' denotes the type B anomaly. So for Einstein gravity
eq.(\ref{EH}), we have
$b_{\frac{d}{2}}=-\frac{2d}{(d!)^2}E_{2n}+...$. Note that
$\sqrt{-G}(f_0+f_1)$ eq.(\ref{f0f1}) is just the Einstein Hilbert
action multiplied by a factor $-\frac{f_0}{2d}$. So
$\sqrt{-G}(f_0+f_1)$ contributes a term $\frac{f_0}{(d!)^2}E_{2n}$
to $b_{\frac{d}{2}}$. This means that the other terms
$\sqrt{-G}(f_2+f_3+...)$ can not contribute to the type A anomaly.

\section{Examples}
In this section, we study some examples to show the application of
our general approach. In particular, we investigate gravity theories
with derivatives of the curvature. Let us recall the main steps of
our approach. We firstly expand the action around a referenced
curvature, and then select the relevant terms with suitable orders.
Finally, we calculate these relevant terms to derive the holographic
Weyl anomaly. This approach can highly decrease the numbers of terms
needed to be computed. For example, there are only two (four)
relevant terms in five (seven) dimensional spacetime, which is just
the number of independent central charges of the corresponding CFTs.

\subsection{Love-Lock Gravity}

Love-Lock gravity is a general theory of gravity whose equations of
motion are only second order in derivatives. The action of Love-Lock
gravity is
\begin{eqnarray}\label{LoveLock}
S&=&\frac{1}{2\kappa^2_{d+1}}\int
d^{d+1}x\sqrt{-\hat{G}}[\frac{d(d-1)}{L^2}+\hat{R}+\sum_{p=2}^{[\frac{d+1}{2}]}c_p
L_{2p}],
\end{eqnarray}
where $L_{2p}$ is defined as
\begin{eqnarray}\label{LoveLockL2p}
L_{2p}=\frac{1}{2^p}\delta^{\nu_1\nu_2...\nu_{2p-1}\nu_{2p}}_{\mu_1\mu_2...\mu_{2p-1}\mu_{2p}}\hat{R}^{\mu_1\mu_2}_{\
\ \ \ \ \nu_1\nu_2}...\hat{R}^{\mu_{2p-1}\mu_{2p}}_{\ \ \ \ \ \ \ \
\ \ \nu_{2p-1}\nu_{2p}}.
\end{eqnarray}
Similar to Einstein gravity, Love-Lock gravity has a well defined
Gibbons-Hawking surface term and Brown-York surface stress
tensor\cite{Dehghani}. There is also an exact form of holographic
entanglement entropy for Love-Lock gravity \cite{Hung,Boer}. Let us
begin to derive the holographic Weyl anomaly for Love-Lock gravity.
For simplicity, we introduce the following notation
\begin{eqnarray}
\lambda_p=(-1)^p\frac{(d-2)!}{(d-2p)!}c_p,\ \ \
f_{\infty}=\frac{L^2}{\tilde{L}^2},
\end{eqnarray}
where $\tilde{L}$ is the curvature scale of the AdS vacua. We set
$\tilde{L}=1$ in this paper. We have assumed that AdS vacua is a
solution to Love-Lock gravity, which yields
\begin{eqnarray}\label{LoveLockf}
1=f_{\infty}-\sum_{p=2}^{[d/2]}\lambda_p (f_{\infty})^p.
\end{eqnarray}

For $d=4$, the action becomes
\begin{eqnarray}\label{4dLoveLock}
S&=&\frac{1}{2\kappa^2_{5}}\int
d^{5}x\sqrt{-\hat{G}}[\frac{12}{L^2}+\hat{R}+\frac{\lambda
L^2}{2}L_4],
\end{eqnarray}
where $L_{4}$ is given by
\begin{eqnarray}\label{4dLoveLockL2p}
L_{4}=\hat{R}_{\mu\nu\rho\sigma}\hat{R}^{\mu\nu\rho\sigma}-4\hat{R}_{\mu\nu}\hat{R}^{\mu\nu}+\hat{R}^2.
\end{eqnarray}
Expanding the action around the referenced curvature
(\ref{referenceR}) and selecting the terms relevant to the Weyl
anomaly, we get
\begin{eqnarray}\label{4dLoveLock1}
S&=&\frac{1}{2\kappa^2_{5}}\int
d^{5}x\sqrt{-\hat{G}}[-\frac{f_0}{8}(\hat{R}+12)+\frac{\lambda
L^2}{2}(\hat{R}-\bar{R})_{\mu\nu\rho\sigma}(\hat{R}-\bar{R})^{\mu\nu\rho\sigma}+...],
\end{eqnarray}
with $f_0=\frac{12}{f_{\infty}}-20+60 \lambda f_{\infty}$. Applying
eq.(\ref{LoveLockf}), we can simplify $f_0$ as $f_0=-8+48\lambda
f_{\infty}$.

Using eqs.(\ref{A},\ref{XX1}), we obtain the Weyl anomaly
(\ref{4dWeyl}) with the charges
\begin{eqnarray}\label{4dLoveLockac}
a=\frac{\pi^2}{\kappa_5^2}(1-6\lambda f_{\infty}), \ \ \
c=\frac{\pi^2}{\kappa_5^2}(1-2\lambda f_{\infty}),
\end{eqnarray}
which exactly agrees with eq.(4.4) of \cite{Hung}.

Similarly, for $d=6$ the action is
\begin{eqnarray}\label{6dLoveLock}
S&=&\frac{1}{2\kappa^2_{7}}\int
d^{7}x\sqrt{-\hat{G}}[\frac{30}{L^2}+\hat{R}+\frac{\lambda
L^2}{12}L_4-\frac{L^4}{24}\mu L_6],
\end{eqnarray}
with
\begin{eqnarray}
L_6=K^3_1-12K^3_2+3K^3_3+16K^3_4-24K^3_5-24K^3_6+4K^3_7+8K^3_8.
\end{eqnarray}
We refer the reader to eq.(\ref{Kni}) for the definitions of
$K^3_i$. Expanding the action around the referenced curvature
(\ref{referenceR}) and selecting the terms relevant to the Weyl
anomaly, we obtain
\begin{eqnarray}\label{6dLoveLock1}
S&=&\frac{1}{2\kappa^2_{7}}\int
d^{7}x\sqrt{-\hat{G}}[-\frac{f_0}{12}(\hat{R}+30)+c^2_1\tilde{K}^2_1+c^3_7\tilde{K}^3_7+c^3_8\tilde{K}^3_8+...],
\end{eqnarray}
with
\begin{eqnarray}
f_0&=&4 (-3 + 10 f_{\infty} \lambda + 45 f_{\infty}^2 \mu),\nonumber\\
c^2_1&=& \frac{f_{\infty}}{12} \lambda + \frac{3}{4} f_{\infty}^2
\mu,\nonumber\\
c^3_7&=&-\frac{f_{\infty}^2}{6} \mu,\nonumber\\
c^3_8&=&-\frac{f_{\infty}^2}{3}\mu.
\end{eqnarray}
It should be mentioned that we have used eq.(\ref{LoveLockf}) to
simplify $f_0$. Applying our formula (\ref{6dAB}), we obtain the
holographic Weyl anomaly (\ref{6dWeylanomaly}) with the
corresponding charges
\begin{eqnarray}
&A&=\frac{\pi^3}{\kappa^2_{7}}\frac{3-10f_{\infty}\lambda-45f_{\infty}^2\mu}{6},\nonumber\\
&B_1&=\frac{1}{\kappa^2_{7}}\frac{-9+26f_{\infty}\lambda+51f_{\infty}^2\mu}{288},\nonumber\\
&B_2&=\frac{1}{\kappa^2_{7}}\frac{-9+34f_{\infty}\lambda+75f_{\infty}^2\mu}{1152},\nonumber\\
&B_3&=\frac{1}{\kappa^2_{7}}\frac{1-2f_{\infty}\lambda-3f_{\infty}^2\mu}{6},\nonumber\\
\end{eqnarray}
which is exactly the same as eq.(5.4) of \cite{Hung}.

To summarize, we have derived the correct holographic Weyl anomaly
for Love-Lock gravity in asymptotically $AdS_5$ and $AdS_7$. It can
be regarded as a test of our general formulas. Our method is much
simpler than the traditional one. First, we make no use of equations
of motion. Second, we only need to calculate a few relevant terms.
It helps a lot to simplify the calculations.

\subsection{$f(R)$ gravity}

Consider $f(\hat{R})$ gravity with the action
\begin{eqnarray}\label{fRderivative1}
S&=&\frac{1}{2\kappa^2_{d+1}}\int d^{d+1}x\sqrt{-\hat{G}}f(\hat{R}),
\end{eqnarray}
which has an asymptotically AdS solution. Expanding the action
around the referenced curvature (\ref{referenceR}), we get
\begin{eqnarray}\label{fRderivative2}
S&=&\frac{1}{2\kappa^2_{d+1}}\int d^{d+1}x\sqrt{-\hat{G}}\big{(} \
f(\bar{R})+f'(\bar{R})(\hat{R}-\bar{R})+\frac{1}{2}f''(\bar{R})(\hat{R}-\bar{R})^2+...\
\big{)}\nonumber\\
&=&\frac{1}{2\kappa^2_{d+1}}\int d^{d+1}x\sqrt{-\hat{G}}\big{(}
-\frac{f(\bar{R})}{2d}(\hat{R}+d^2-d)+\frac{1}{2}f''(\bar{R})(\hat{R}-\bar{R})^2+...\
\big{)}
\end{eqnarray}
Note that $(\hat{R}-\bar{R})^2\sim o(\rho^4)$ does not contribute to
the Weyl anomaly for $d=4,6$. Thus we only need to calculate the
first term of the second line of the above equation. And
$f(\hat{R})$ gravity behaves effectively as Einstein gravity with a
negative cosmological constant for $d=4,6$, just replacing
$\frac{1}{2\kappa^2_{d+1}}$ by
$-\frac{f(\bar{R})}{2d}\frac{1}{2\kappa^2_{d+1}}$. It is consistent
with the fact that $f(\hat{R})$ gravity is equivalent to Einstein
gravity plus a scalar field. Now it is clear that the Weyl Anomaly
of $f(\hat{R})$ gravity is just $-\frac{f(\bar{R})}{2d}$ times the
one of Einstein gravity for $d=4,6$.

\subsection{Critical Gravity}

According to \cite{HongLu}, the one-parameter critical theory is
given by the action
\begin{eqnarray}\label{criticalaction}
S&=&\frac{1}{2\kappa^2_{d+1}}\int
d^{d+1}x\sqrt{-\hat{G}}[\hat{R}+\frac{d(d-1)}{L^2}-\frac{L^2}{4(d-2)}
\hat{C}^{\mu\nu\rho\sigma}\hat{C}_{\mu\nu\rho\sigma}].
\end{eqnarray}
Here $\hat{C}_{\mu\nu\rho\sigma}$ is the Weyl tensor and
\begin{eqnarray}\label{criticalWeyltensor}
\hat{C}^{\mu\nu\rho\sigma}\hat{C}_{\mu\nu\rho\sigma}=\hat{R}^{\mu\nu\rho\sigma}\hat{R}_{\mu\nu\rho\sigma}-\frac{4}{d-1}\hat{R}^{\mu\nu}\hat{R}_{\mu\nu}+\frac{2}{d(d-1)}\hat{R}^2.
\end{eqnarray}
For simplicity, we set $L=1$ below. This critical gravity has a
unique AdS vacuum in which there are only massless spin-2 modes.
Besides, the mass and angular momenta of all asymptotically Kerr-AdS
and Schwarzschild-AdS black holes vanish.

Expanding the action around the referenced curvature
(\ref{referenceR}) and keeping only the relevant terms, we get
\begin{eqnarray}\label{criticalaction1}
S&=&\frac{1}{2\kappa^2_{d+1}}\int
d^{d+1}x\sqrt{-\hat{G}}[\hat{R}+d(d-1)-\frac{1}{4(d-2)}
\tilde{K}^2_1+...].
\end{eqnarray}
It is easy to observe that $f_0=-2d$ and $c^2_1=-\frac{1}{4(d-2)}$.
Applying the general formulas (\ref{ac},\ref{6dAB}), we can easily
obtain the holographic Weyl anomaly. For $d=4$, we get the
holographic Weyl anomaly (\ref{4dWeyl}) with the charges
\begin{eqnarray}\label{criticalac}
a=\frac{\pi^2}{\kappa_5^2}, \ \ \ c=0.
\end{eqnarray}
And for $d=6$, we obtain the holographic Weyl anomaly
(\ref{6dWeylanomaly}) with the charges
\begin{eqnarray}\label{6dAB}
&&A=\frac{\pi^3}{2\kappa^2_{7}},\ \ \ \ \ \ B_1=-\frac{1}{48\kappa^2_{7}},\nonumber\\
&&B_2=-\frac{1}{96\kappa^2_{7}},\ B_3=0.
\end{eqnarray}

As we mentioned above, the mass of the black holes in critical
gravity vanish. One may doubt that the critical gravity is a trivial
theory. However, as we have derived above, the central charges of
the CFT dual to critical gravity is nonzero generally, which implies
that the critical gravity is indeed non-trivial. It is also
interesting to note that some of the type B anomaly vanish for the
critical gravity.

\subsection{Gravity with derivatives of curvatures}

The general method developed in Sect.2 can be easily generalized to
the modified gravity with derivatives of curvatures. We firstly
expand the action around the referenced curvature
$\bar{R}_{\mu\nu\rho\rho}$ (\ref{referenceR}), then select and
calculate the terms which contribute to the Weyl Anomaly. Let us
study an example with the action
\begin{eqnarray}\label{actionderivative}
S&=&\frac{1}{2\kappa^2_{d+1}}\int
d^{d+1}x\sqrt{-\hat{G}}(\hat{R}+\frac{d^2-d}{L^2}+\lambda_1\hat{R}\Box\hat{R}+\lambda_2\hat{R}_{\mu\nu}\Box\hat{R}^{\mu\nu}+\lambda_3\hat{R}_{\mu\nu\rho\sigma}\Box\hat{R}^{\mu\nu\rho\sigma}).\nonumber\\
\end{eqnarray}
The first two terms above are just Einstein-Hilbert action with a
negative cosmological constant. We have calculated the Weyl anomaly
from these terms. Expanding the last three terms around the
referenced curvature, we have
\begin{eqnarray}\label{actionderivative1}
S&=&\frac{1}{2\kappa^2_{d+1}}\int
d^{d+1}x\sqrt{-\hat{G}}(\hat{R}+\frac{d^2-d}{L^2}\nonumber\\
&+&\lambda_1(\hat{R}-\bar{R})\Box(\hat{R}-\bar{R})+\lambda_2(\hat{R}-\bar{R})_{\mu\nu}\Box(\hat{R}-\bar{R})^{\mu\nu}+\lambda_3(\hat{R}-\bar{R})_{\mu\nu\rho\sigma}\Box(\hat{R}-\bar{R})^{\mu\nu\rho\sigma}\
).\nonumber\\
\end{eqnarray}
Here we have dropped some total derivatives. Applying $\Box\sim
o(1)$ together with eqs.(\ref{order1},\ref{order2}), we obtain
\begin{eqnarray}\label{actionderivative1}
&&\sqrt{-\hat{G}}(\hat{R}-\bar{R})\Box(\hat{R}-\bar{R})\sim
o(\rho^{3-\frac{d}{2}})\nonumber\\
&&\sqrt{-\hat{G}}(\hat{R}-\bar{R})_{\mu\nu}\Box(\hat{R}-\bar{R})^{\mu\nu}\sim o(\rho^{3-\frac{d}{2}})\nonumber\\
&&\sqrt{-\hat{G}}(\hat{R}-\bar{R})_{\mu\nu\rho\sigma}\Box(\hat{R}-\bar{R})^{\mu\nu\rho\sigma}\sim
o(\rho^{1-\frac{d}{2}}).\nonumber\\
\end{eqnarray}
For $(d=4,\ 6)$, it is clear that only the last term contributes to
the Weyl anomaly. For simplicity, let us denote
\begin{eqnarray}
&&K^3_i=(\hat{R}^3,\hat{R}\hat{R}_{\mu\nu}\hat{R}^{\mu\nu},\hat{R}\hat{R}_{\mu\nu\rho\sigma}\hat{R}^{\mu\nu\rho\sigma},\hat{R}_{\mu}^{\nu}\hat{R}_{\nu}^{\rho}\hat{R}_{\rho}^{\mu},
\hat{R}^{\mu\nu}\hat{R}^{\rho \sigma}\hat{R}_{\mu \rho \sigma\nu},
\hat{R}_{\mu \nu}\hat{R}^{\mu \rho \sigma \lambda}\hat{R}^{\nu}_{\ \rho \sigma \lambda},\nonumber\\
&&\ \ \ \ \ \ \ \ \hat{R}_{\mu \nu \rho \sigma}\hat{R}^{\mu \nu
\lambda\chi}\hat{R}^{\rho \sigma}_{\ \ \lambda\chi},\hat{R}_{\nu \nu
\rho\sigma}\hat{R}^{\nu\lambda\chi\sigma}\hat{R}^{\nu\ \ \rho}_{\
\lambda\chi},\ \hat{R}\Box\hat{R},\
\hat{R}_{\mu\nu}\Box\hat{R}^{\mu\nu},\
\hat{R}_{\mu\nu\rho\sigma}\Box\hat{R}^{\mu\nu\rho\sigma}).
\end{eqnarray}
Then the last term becomes
$\sqrt{-\hat{G}}K^3_{11}=\sqrt{-\hat{G}}\tilde{K}^3_{11}$, where
$\tilde{K}=K|_{\hat{R}\rightarrow \hat{R}-\bar{R}}$. Note that
\begin{eqnarray}\label{totalderivative}
K^3_{11}+K^3_9-4K^3_{10}+4(K^3_4+K^3_5)-2K^3_6+K^3_7-4K^3_8=\nabla_{\mu}J^{\mu}
\end{eqnarray}
is a total derivative. Since total derivatives do not contribute to
the anomaly, this formula can help us to rewrite $\tilde{K}^3_{11}$
in terms of the other $\tilde{K}^n_i$ ($n\leq 3,i\leq 10$). We can
derive
\begin{eqnarray}\label{6dK11r}
\sqrt{-\hat{G}}\tilde{K}^3_{11}&=&\sqrt{-\hat{G}}(\hat{R}-\bar{R})_{\mu\nu\rho\sigma}\Box(\hat{R}-\bar{R})^{\mu\nu\rho\sigma},\nonumber\\
&=&\sqrt{-\hat{G}}\hat{R}_{\mu\nu\rho\sigma}\Box\hat{R}^{\mu\nu\rho\sigma}+...\nonumber\\
&=&\sqrt{-\hat{G}}(-4K^3_4-4K^3_5+2K^3_6-K^3_7+4K^3_8-K^3_9+4K^3_{10})+...\nonumber\\
&=&\sqrt{-\hat{G}}(-4\tilde{K}^3_4-4\tilde{K}^3_5+2\tilde{K}^3_6-\tilde{K}^3_7+4\tilde{K}^3_8-\tilde{K}^3_9+4\tilde{K}^3_{10}\nonumber\\
&& \ \ \ \ \ \ \ \ \
-4\tilde{K}^2_3+4(2+d)\tilde{K}^2_2-2d\tilde{K}^2_1+...  )
\end{eqnarray}
where ``...'' denotes the total derivatives.

For $d=4$, it is clear that only $\tilde{K}^2_1$ in
eq.(\ref{6dK11r}) contributes to the Weyl anomaly. Using
eq.(\ref{XX1}), we have
\begin{eqnarray}\label{4dderivative}
\sqrt{-\hat{G}}\tilde{K}^3_{11}&=&\sqrt{-\hat{G}}(-2d\tilde{K}^2_{1}+...)\nonumber\\
&=&\frac{\sqrt{-g_{(0)}}}{2\rho}(-2dC_{ijkl}C^{ijkl}+...).
\end{eqnarray}
Thus we obtain the 4d Weyl anomaly (\ref{4dWeyl}) with charges
\begin{eqnarray}\label{acderivative}
a=\frac{\pi^2}{\kappa_5^2}, \ \ \
c=(1-64\lambda_3)\frac{\pi^2}{\kappa_5^2}.
\end{eqnarray}

For $d=6$, only terms $\tilde{K}^2_1,\tilde{K}^3_7,\tilde{K}^3_8$ in
eq.(\ref{6dK11r}) contribute to the Weyl anomaly. Using
eqs.(\ref{6df2},\ref{6dK37}), we can derive
\begin{eqnarray}\label{6dK11}
\sqrt{-\hat{G}}\tilde{K}^3_{11}&=&\sqrt{-\hat{G}}(-\tilde{K}^3_7+4\tilde{K}^3_8-2d\tilde{K}^2_{1}+...)\nonumber\\
&=&\frac{\sqrt{-\hat{g_{(0)}}}}{2\rho}(8I_1-2I_2-I_3)+...
\end{eqnarray}
Then we obtain the 6d Weyl anomaly (\ref{6dWeylanomaly}) with
charges
\begin{eqnarray}\label{6dAB}
&&A=\frac{\pi^3}{2\kappa^2_{7}},\nonumber\\
&&B_1=\frac{1}{2\kappa^2_{7}}(-\frac{1}{16}+8\lambda_3),\nonumber\\
&&B_2=\frac{1}{2\kappa^2_{7}}(-\frac{1}{64}-2\lambda_3),\nonumber\\
&&B_3=\frac{1}{2\kappa^2_{7}}(\frac{1}{192}-\lambda_3).
\end{eqnarray}

Interestingly, although terms ($\hat{R}\Box\hat{R},\
\hat{R}_{\mu\nu}\Box\hat{R}^{\mu\nu}$) in the action affect the
equations of motion, they do not contribute to the holographic Weyl
anomaly. This is a reflection of the fact that the holographic Weyl
anomaly for $d=4,6$ is independent of the equations of motion.

\section{Holographic Entanglement Entropy}

In this section, we propose a formula of the holographic
entanglement entropy in asymptotically $AdS_5$ and compare it with universal term of entanglement entropy for 4d CFTs\cite{Solodukhin} . For simplicity, we work in the Euclidean signature. So the entropy formula is different from the usual Lorentzian one by a minus sign. Based on the works of Myers \cite{Hung} and Fursaev et al \cite{Solodukhin1}, we assume
the holographic entanglement of general higher derivative gravity in asymptotically $AdS_5$ is
\begin{eqnarray}\label{HE}
S_{HE}=-\frac{2\pi}{\kappa_{5}^2}\int_m d^3x\sqrt{h}\frac{\delta
f}{\delta
\hat{R}^{\mu\nu\rho\sigma}}(n^{\mu}n^{\rho})(n^{\nu}n^{\sigma})+S_{K},
\end{eqnarray}
where $n^{\mu}_{\iota}$ with $\iota=1,2$ are the two vectors
orthogonal to $m$ and ($n^{\mu}n^{\rho}$) denotes
$n^{\mu}_{\iota}n^{\rho \iota}$. The first term is just the Wald
entropy while the second term $S_K$ denotes the contribution from
the extrinsic curvature
\begin{eqnarray}\label{K}
S_{K}=\frac{4\pi}{\kappa_{5}^2}\int_m d^3x\sqrt{h} (c^2_1
K^{\iota\alpha}_{\ \beta}K_{\iota\beta}^{\ \alpha}+\frac{1}{4}c^2_2
K^{\iota \alpha}_{\alpha}K_{\iota \beta}^{\beta}).
\end{eqnarray}
See eq.(\ref{Expandvector}) for the definition of $c^n_i$. Note that
$S_{K}$ is designed to be consistent with the formula of the
holographic entanglement entropy of Love-Lock gravity
\cite{Boer,Hung} and the curvature-squared gravity
\cite{Solodukhin1}.

The universal logarithmic term of the entanglement entropy for 4d
CFTs is
\begin{eqnarray}\label{EE}
S_{EE}=\log(l/\delta)\frac{1}{2\pi}\int_{\Sigma}
d^2x\sqrt{h}[aR_{\Sigma}-c(C^{abcd}h_{ac}h_{bd}-K^{\iota ab}K_{\iota
ab}+\frac{1}{2}K^{\iota a}_aK_{\iota b}^b) ],
\end{eqnarray}
which was found in \cite{Solodukhin} using the conformal symmetry
and the holography. We compare this formula with our proposal
(\ref{HE}) below.

Now let us derive the universal contribution to the entanglement
entropy from eq.(\ref{HE}). Firstly, we focus on the Wald entropy
term. Applying the methods of \cite{Theisen,Hung}, we can derive the
logarithmic term as
\begin{eqnarray}\label{Wald}
S_{W}=\frac{\log(l/\delta)}{2\pi}\int_{\Sigma}
d^2x\sqrt{h}[a(R_{\Sigma}+K^{\iota ab}K_{\iota
ab}-\frac{1}{2}K^{\iota a}_aK_{\iota b}^b) -cC^{abcd}h_{ac}h_{bd}].
\end{eqnarray}
A useful technique in the above derivation is that we expand $S_W$
around $\bar{R}_{\mu\nu\rho\sigma}$ eq.(\ref{referenceR}):
\begin{eqnarray}\label{expand}
\frac{\delta f}{\delta
\hat{R}^{\mu\nu\rho\sigma}}(n^{\mu}n^{\rho})(n^{\nu}n^{\sigma})&=&P_{\mu\nu\rho\sigma}|_{\bar{R}}(n^{\mu}n^{\rho})(n^{\nu}n^{\sigma})\nonumber\\
&+&\frac{\delta^2f}{\delta\hat{R}_{\mu\nu\rho\sigma}\delta\hat{R}_{\mu_1\nu_1\rho_1\sigma_1}}|_{\bar{R}}(\hat{R}_{\mu_1\nu_1\rho_1\sigma_1}-\bar{R}_{\mu_1\nu_1\rho_1\sigma_1})(n^{\mu}n^{\rho})(n^{\nu}n^{\sigma})\nonumber\\
&+&o(\rho^2)\nonumber\\
&=&-\frac{f_0}{2d}+n_1(\hat{R}_{\mu\nu\rho\sigma}-\bar{R}_{\mu\nu\rho\sigma})(n^{\mu}n^{\rho})(n^{\nu}n^{\sigma})+o(\rho^2)\nonumber\\
&=&-\frac{f_0}{2d}+\rho \ n_1 C^{abcd}h_{ac}h_{bd}+o(\rho^2).
\end{eqnarray}
In the above derivations, we have used the following useful formulas
\begin{eqnarray}\label{formulas}
&&\hat{R}_{\mu\nu\rho\sigma}(n^{\mu}n^{\rho})(n^{\nu}n^{\sigma})=-2+\rho
C^{abcd}h_{ac}h_{bd}+o(\rho^2),\nonumber\\
&&\hat{R}_{\mu\nu}(n^{\mu}n^{\nu})=-8+o(\rho^2),\ \ \
\hat{R}=-20+o(\rho^2).
\end{eqnarray}

Let us go on to calculate $S_K$. After some calculations, we obtain
\begin{eqnarray}\label{SSK}
S_K&=&\frac{4\pi}{\kappa_{5}^2}\int_m d^3x\sqrt{h} (c^2_1
K^{\iota\alpha}_{\ \beta}K_{\iota\beta}^{\ \alpha}+\frac{1}{4}c^2_2
K^{\iota \alpha}_{\alpha}K_{\iota \beta}^{\beta})\nonumber\\
&=&(c-a)\frac{\log(l/\delta)}{2\pi}\int_{\Sigma} d^2x\sqrt{h}[
K^{\iota ab}K_{\iota ab}-\frac{1}{2}K^{\iota a}_aK_{\iota b}^b].
\end{eqnarray}
From eqs.(\ref{Wald},\ref{SSK}), we finally obtain the logarithmic
term of $S_{HE}$ as
\begin{eqnarray}\label{HHE}
S_{EE}=\log(l/\delta)\frac{1}{2\pi}\int_{\Sigma}
d^2x\sqrt{h}[aR_{\Sigma} -c(C^{abcd}h_{ac}h_{bd}-K^{\iota ab}K_{\iota
ab}+\frac{1}{2}K^{\iota a}_aK_{\iota b}^b)],
\end{eqnarray}
which agrees with the result of CFTs eq.(\ref{EE}).

Let us comment on our results. First, the
holographic entanglement entropy takes the same form as the Wald
entropy for gravity theories with the same `a' charge and `c' charge
(such as Einstein gravity and $f(R)$ gravity). Second, our proposal of
the holographic entanglement entropy (\ref{HE}) only works effectively in asymptotically $AdS_5$.
By ``effectively'', we mean that it is the leading term of holographic entanglement entropy in asymptotically $AdS_5$. And the correct formula of  holographic entanglement entropy must reduce to ours in asymptotically $AdS_5$. After the work is finished, other interesting formulas of holographic entanglement entropy for higher derivative gravity are proposed by Dong \cite{Dong} and Camps \cite{Camps}, respectively.  Camps later realized that his proposal only applies to curvature-squared gravity and can be regarded as a special case of Dong's proposal. Thus we focus on Dong's proposal below. We shall show that our formula is just the leading term of Dong's proposal in asymptotically $AdS_5$. Since the higher order terms do not contribute to the universal term of entanglement entropy for 4d CFTs, we actually shall prove Dong's proposal yields the correct universal term of entanglement entropy for 4d CFTs. 

According to Dong\cite{Dong}, the general $S_K$ in (\ref{HE}) should be
\begin{eqnarray}\label{KDong}
S_{K}=&&\frac{\pi}{\kappa_{5}^2}\int_m d^3x\sqrt{h} \big{(} \frac{\partial^2 f}{\partial R_{\mu_1\rho_1\nu_1\sigma_1}\partial R_{\mu_2\rho_2\nu_2\sigma_2}} \big{)}_{\alpha} \frac{2K_{\lambda_1\rho_1\sigma_1}K_{\lambda_2\rho_2\sigma_2}}{q_{\alpha}+1} \nonumber\\
&&\times[ (n_{\mu_1\mu_2}n_{\nu_1\nu_2}-\varepsilon_{\mu_1\mu_2}\varepsilon_{\nu_1\nu_2})n^{\lambda_1\lambda_2}+ (n_{\mu_1\mu_2}\varepsilon_{\nu_1\nu_2}+\varepsilon_{\mu_1\mu_2}n_{\nu_1\nu_2})\varepsilon^{\lambda_1\lambda_2}].
\end{eqnarray}
 Please refer to \cite{Dong} for the definitions of $q_{\alpha}, n_{\mu\nu}, \varepsilon_{\mu\nu} $. In asymptotically $AdS_5$, using the method developed in this paper, we can expand the above $S_K$ 
around $\bar{R}_{\mu\nu\rho\sigma}$ (\ref{referenceR}). Let us focus on the leading term. From eq.(\ref{Expandvector}), we have
\begin{eqnarray}\label{XXX}
\frac{\partial^2 f}{\partial R_{\mu_1\rho_1\nu_1\sigma_1}\partial R_{\mu_2\rho_2\nu_2\sigma_2}}|_{\bar{R}}=2(c^2_1X^2_1+c^2_2X^2_2+c^2_3X^2_3)^{\mu_1\rho_1\nu_1\sigma_1\mu_2\rho_2\nu_2\sigma_2},
\end{eqnarray}
where the tensor $X^2_i$ is defined in eqs.(\ref{X1}-\ref{X3}). Since $X^2_i$ contains only the metric, according to \cite{Dong} we have $q_{\alpha}=0$.  Thus, the leading term of eq.(\ref{KDong}) in asymptotically $AdS_5$ becomes
\begin{eqnarray}\label{LeadingK}
S_{K}&=&\frac{4\pi}{\kappa_{5}^2}\int_m d^3x\sqrt{h} (c^2_1X^2_1+c^2_2X^2_2+c^2_3X^2_3)^{\mu_1\rho_1\nu_1\sigma_1\mu_2\rho_2\nu_2\sigma_2} K_{\lambda_1\rho_1\sigma_1}K_{\lambda_2\rho_2\sigma_2} \nonumber\\
&&\ \ \times[ (n_{\mu_1\mu_2}n_{\nu_1\nu_2}-\varepsilon_{\mu_1\mu_2}\varepsilon_{\nu_1\nu_2})n^{\lambda_1\lambda_2}+ (n_{\mu_1\mu_2}\varepsilon_{\nu_1\nu_2}+\varepsilon_{\mu_1\mu_2}n_{\nu_1\nu_2})\varepsilon^{\lambda_1\lambda_2}]+...\nonumber\\
&=&\frac{4\pi}{\kappa_{5}^2}\int_m d^3x\sqrt{h} (c^2_1
K^{\iota\alpha}_{\ \beta}K_{\iota\beta}^{\ \alpha}+\frac{1}{4}c^2_2
K^{\iota \alpha}_{\alpha}K_{\iota \beta}^{\beta})+...
\end{eqnarray}
which is exactly our proposal eq.(\ref{K}). Here ``...'' denotes the high order terms in $\rho$ and these terms do not contribute to the universal log term of entanglement entropy. Now we have proved that our formula eq.(\ref{HE}) is the leading term of Dong's proposal in asymptotically $AdS_5$. Since only the leading term contributes to universal log term, we actually prove that Dong's proposal yields the correct universal term of entanglement entropy for 4d CFTs. This is a nontrivial test of Dong's proposal.

\section{Conclusions}

We develop a simple approach to derive the holographic Weyl anomaly
from general higher derivative gravity. Applying our approach, we
only need to calculate a few relevant terms which highly simplify the
derivations of the Weyl anomaly. It is remarkable that we make no
use of equations of motion to derive all the central charges of 4d
and 6d CFTs. As an application of our results, we propose a formula
of holographic entanglement entropy for general higher derivative
gravity in asymptotic $AdS_5$. Our proposal is consistent with the
holographic entanglement entropy of Einstein gravity and Love-Lock
gravity. Furthermore, our proposal can yield the correct universal
term of entanglement entropy for 4d CFT. We find that our formula of of holographic entanglement entropy is just the leading term of Dong's proposal in asymptotic $AdS_5$.  Since only the leading term contributes to universal log term of entanglement entropy, we actually prove that Dong's proposal can yield the correct universal term of entanglement entropy for 4d CFTs. This is a nontrivial test of Dong's proposal. It is interesting to check if Dong's proposal could yield the universal term of entanglement entropy for 6d CFTs \cite{Zhong}.
 We hope to address this problem in future.

\section*{Acknowledgements}

We thank Wei Li and S. Theisen for helpful discussions. We also
thank Albert Einstein Institute for hospitality.

\appendix

\section{Useful formulas}

In this appendix, we provide some useful formulas of Riemann tensor.
Our conventions for the curvature tensors are
$[\nabla_{\mu},\nabla_{\nu}]V^{\rho}=R_{\mu\nu\ \ \sigma}^{\ \ \
\rho}V^{\sigma}, R_{\mu\nu}=R^{\rho}_{\ \mu\rho\nu}$. We assume the
metric in the bulk takes the following form:
\begin{eqnarray}\label{mmetric}
ds^2=\hat{G}_{\mu\nu}dx^{\mu}dx^{\nu}=\frac{1}{4\rho^2}d\rho^2+\frac{1}{\rho}g_{ij}dx^idx^j.
\end{eqnarray}
According to \cite{Nojiri}, we have formulas for scalar curvature

\begin{eqnarray} \label{iii}
\hat R&=&-{d^2 + d \over l^2}+\rho R +{2(d-1) \rho \over
l^2}g^{ij}g'_{ij} + {3\rho^2 \over l^2} g^{ij}g^{kl}g'_{ik}g'_{jl}
\nonumber\\ && - {4\rho^2 \over l^2}g^{ij}g''_{ij} - {\rho^2 \over
l^2}g^{ij}g^{kl}g'_{ij}g'_{kl}
\end{eqnarray}
for Ricci tensor \begin{eqnarray} \label{iv} \hat R_{\rho\rho}&=&-{d
\over 4\rho^2} - {1 \over 2}g^{ij}g''_{ij} + {1 \over
4}g^{ik}g^{lj}g'_{kl}g'_{ij} \nonumber\\
\hat
R_{i\rho}&=&\frac{1}{2}g^{jk}(g'_{ki;j}-g'_{kj;i})\nonumber\\&=&{1
\over 2}g^{jk}g'_{ki,j} - {1 \over 2}g^{kj}g'_{jk,i} + {1 \over
2}g^{jk}_{,j}g'_{ki}  + {1 \over 4}g^{kl}g'_{li}g^{jm}g_{jm,k} - {1
\over 4}g^{kj}_{,i}g'_{jk}\nonumber\\
\hat R_{ij}&=&R_{ij}\nonumber\\ && - {2\rho \over l^2}g''_{ij} +
{2\rho \over l^2}g^{kl}g'_{ki}g'_{lj} - {\rho \over
l^2}g'_{ij}g^{kl}g'_{kl}  - {2-d \over l^2}g'_{ij} + {1 \over
l^2}g_{ij}g^{kl}g'_{kl} - {d \over l^2\rho}g_{ij}
\end{eqnarray}

and for the Riemann tensor
\begin{eqnarray}\label{v}
\hat R_{i\rho j\rho}&=& - {1 \over 4\rho^3}g_{ij} + {1 \over
4\rho}g^{kl}g'_{ki}g'_{lj} - {1 \over 2\rho}g''_{ij} \nonumber \\
\hat R_{\rho ijk}&=&\frac{1}{2\rho}(g'_{ij;k}-g'_{ik;j})\nonumber\\
&=&{1 \over 4\rho }\left\{ 2g'_{ij,k} - 2g'_{ik,j} -
g^{lm}\left(g_{im,k} + g_{km,i} - g_{ik,m}\right)g'_{lj} \right.
\nonumber\\ && \left. + g^{lm}\left(g_{im,j} + g_{jm,i} -
g_{ij,m}\right)g'_{lk} \right\}
\nonumber\\
\hat R_{ijkl}&=&{1 \over \rho}R_{ijkl}\nonumber\\&& - {1 \over
\rho^2 l^2}\left\{\left( g_{jl}-\rho g'_{jl}\right) \left(g_{ik} -
\rho g'_{ik}\right) -\left( g_{jk}-\rho g'_{jk}\right) \left(g_{il}
- \rho g'_{il}\right)\right\}.
\end{eqnarray}
Here `` $'$ " denotes the derivative with respect to $\rho$. ``$;$ "
and ``$R$'' stand for the the covariant derivative and curvature
with respect to $g_{ij}$, respectively.

Let us define a reference curvature as
\begin{eqnarray}\label{referenceR}
\bar{R}=-d(d+1),\ \ \bar{R}_{\mu\nu}=-d \hat{G}_{\mu\nu},\ \
\bar{R}_{\mu\nu\rho\sigma}=-(\hat{G}_{\mu\rho}\hat{G}_{\nu\sigma}-\hat{G}_{\mu\sigma}\hat{G}_{\nu\rho}).
\end{eqnarray}
Please do not confuse the reference curvature
$\bar{R}_{\mu\nu\rho\sigma}$ with
$\hat{R}_{\mu\nu\rho\sigma}|_{AdS}=-(\hat{G}_{(0)\mu\rho}\hat{G}_{(0)\nu\sigma}-\hat{G}_{(0)\mu\sigma}\hat{G}_{(0)\nu\rho})$.
Using eqs.(\ref{iii},\ref{iv},\ref{v}), we observe that
\begin{eqnarray}\label{order1}
\hat{R}-\bar{R}=o(\rho),\ \ \hat{R}_{\mu\nu}-\bar{R}_{\mu\nu}=o(1),\
\ \
\hat{R}_{\mu\nu\rho\sigma}-\bar{R}_{\mu\nu\rho\sigma}=o(\frac{1}{\rho}).
\end{eqnarray}
Applying eq.(\ref{g1}), we can get stronger conditions
\begin{eqnarray}\label{order2}
\hat{R}-\bar{R}=o(\rho^2),\ \ \hat{R}_{ij}-\bar{R}_{ij}=o(\rho),\ \
\ \ \hat{R}_{i\rho}-\bar{R}_{i\rho}=o(\rho).
\end{eqnarray}
In fact, in general, we have
\begin{eqnarray}\label{order3}
f(\hat{R}_{\mu\nu\rho\sigma})-f(\bar{R}_{\mu\nu\rho\sigma})=o(\rho^2),
\end{eqnarray}
where $f$ is a scalar function. The above equations can be used to
simplify the calculations of Weyl anomaly.

\section{Coefficients}

In this appendix, we provide a general method to derive the
coefficients $c^n_i$ of eq.(\ref{Expandvector}):
\begin{eqnarray}\label{Expandvector1}
&&\frac{1}{n!}\frac{\delta^{n}f}{\delta
\hat{R}^{\mu_1\nu_1\rho_1\sigma_1}...\delta
\hat{R}^{\mu_{n}\nu_{n}\rho_{n}\sigma_{n}}}|_{\bar{R}}=\sum_{i=1}^{m}c^n_i
X^{n}_{\ i \
\mu_1\nu_1\rho_1\sigma_1,...,\mu_{n}\nu_{n}\rho_{n}\sigma_{n}},\nonumber\\
&&X^n_{\ i \
\mu_1\nu_1\rho_1\sigma_1,...,\mu_{n}\nu_{n}\rho_{n}\sigma_{n}}=\frac{1}{n!}\frac{\delta^{n}K^n_i}{\delta
\hat{R}^{\mu_1\nu_1\rho_1\sigma_1}...\delta
\hat{R}^{\mu_{n}\nu_{n}\rho_{n}\sigma_{n}}}.
\end{eqnarray} For simplicity, we
focus on the case that $f(\hat{R}_{\mu\nu\rho\sigma})$ contains only
the curvature but not the derivatives of the curvature. To get
$c^n_i$, we need to find a class of tensors $Y^n_j$ with the
conditions:
\begin{eqnarray}\label{newvectorY}
&&Y^{n}_{\ j \
\mu_1\nu_1\rho_1\sigma_1,...,\mu_{n}\nu_{n}\rho_{n}\sigma_{n}}=\sum_{i=1}^{m}(x^n_j)_i
X^{n}_{\ i \
\mu_1\nu_1\rho_1\sigma_1,...,\mu_{n}\nu_{n}\rho_{n}\sigma_{n}},\nonumber\\
&&Y^{n}_{\ j}\ast X^{n}_{\ i}=Y^{n}_{\ j \
\mu_1\nu_1\rho_1\sigma_1,...,\mu_{n}\nu_{n}\rho_{n}\sigma_{n}}X_{\
i}^{n\
\mu_1\nu_1\rho_1\sigma_1,...,\mu_{n}\nu_{n}\rho_{n}\sigma_{n}}=\delta_{ij}.
\end{eqnarray}
Then one can obtain $c^n_i$ as
\begin{eqnarray}\label{oefficient}
c^n_i=\frac{1}{n!}Y^{n}_i\ast\frac{\delta^n f}{(\delta
\hat{R})^n}|_{\hat{R}\rightarrow\bar{R}}.
\end{eqnarray}
Note that the solution to $Y^{n}_i$ is unique. In general, the
calculation is highly non-trivial. We list the results for $d=4$ and
$d=6$ below.

For $d=4$, only $c^2_1$ is relevant to the Weyl anomaly. Solving
eq.(\ref{newvectorY}), we get
\begin{eqnarray}\label{4dY}
&&c^2_1=\frac{1}{2}Y^{2}_1\ast\frac{\delta^2 f}{(\delta
\hat{R})^2}|_{\hat{R}\rightarrow\bar{R}}\nonumber\\
&&Y^{2}_{1}=\frac{12}{(d+3)(d^3+d^2-4d-4)}X^2_1-\frac{48}{(d^2-1)(d+3)(d^2-4)}X^2_2+\frac{24}{d(d^2-1)(d+3)(d^2-4)}X^2_3. \nonumber\\
\end{eqnarray}
Set $d=4$, we obtain eq.(\ref{Y}).

For $d=6$, only $c^2_1, c^3_7, c^3_8$ contribute to the Weyl
anomaly. We have derived $c^2_1$ as above. For $c^3_7$, we have
\begin{eqnarray}\label{6dY7}
&&c^3_7=\frac{1}{3!}Y^{3}_7\ast\frac{\delta^3 f}{(\delta
\hat{R})^3}|_{\hat{R}\rightarrow\bar{R}}\nonumber\\
&&Y^{3}_{7}=\sum_{i=1}^8(x^3_7)_i X^3_i,\nonumber\\
\end{eqnarray}
where $(x^3_7)_i$ are given by
\begin{eqnarray}\label{x37}
&&(x^3_7)_1=\frac{64 (48 + 65 d + d^2)}{d (1 + d)^2 (2880 - 2304 d -
1796 d^2 + 976 d^3 + 389 d^4 - 116 d^5 -
   34 d^6 + 4 d^7 + d^8)},\nonumber\\
&&(x^3_7)_2=-\frac{192 (48 + 65 d + d^2)}{(1 + d)^2 (2880 - 2304 d -
1796 d^2 + 976 d^3 + 389 d^4 - 116 d^5 -
   34 d^6 + 4 d^7 + d^8)},\nonumber\\
&&(x^3_7)_3=\frac{288 (24 - 4 d - 7 d^2 + 5 d^3)}{d (1 + d)^2 (2880
- 2304 d - 1796 d^2 + 976 d^3 + 389 d^4 - 116 d^5 -
   34 d^6 + 4 d^7 + d^8)},\nonumber\\
&&(x^3_7)_4=\frac{128 (144 + 24 d + 23 d^2 + 31 d^3)}{d (1 + d)^2
(2880 - 2304 d - 1796 d^2 + 976 d^3 + 389 d^4 - 116 d^5 -
   34 d^6 + 4 d^7 + d^8)},\nonumber\\
&&(x^3_7)_5=-\frac{192 (-144 - 24 d + 25 d^2 + 34 d^3 + d^4)}{d (1 +
d)^2 (2880 - 2304 d - 1796 d^2 + 976 d^3 + 389 d^4 - 116 d^5 -
   34 d^6 + 4 d^7 + d^8)},\nonumber\\
&&(x^3_7)_6=-\frac{576 (24 - 4 d - 7 d^2 + 5 d^3)}{(1 + d)^2 (2880 -
2304 d - 1796 d^2 + 976 d^3 + 389 d^4 - 116 d^5 -
   34 d^6 + 4 d^7 + d^8)},\nonumber\\
&&(x^3_7)_7=\frac{16 (-144 + 192 d + 109 d^2 - 24 d^3 - 39 d^4 + 14
d^5)}{d (1 + d)^2 (2880 - 2304 d - 1796 d^2 + 976 d^3 + 389 d^4 -
116 d^5 -
   34 d^6 + 4 d^7 + d^8)},\nonumber\\
&&(x^3_7)_8=\frac{64 (144 - 264 d - 25 d^2 + 33 d^3 + 3 d^4 +
d^5)}{(-3 + d) (-1 + d) d (1 + d)^2 (3 + d) (5 + d) (-16 + d^2) (-4
+ d^2)}.
\end{eqnarray}

Similarly, for $c^3_7$, we have
\begin{eqnarray}\label{6dY8}
&&c^3_8=\frac{1}{3!}Y^{3}_8\ast\frac{\delta^3 f}{(\delta
\hat{R})^3}|_{\hat{R}\rightarrow\bar{R}}\nonumber\\
&&Y^{3}_{8}=\sum_{i=1}^8(x^3_8)_i X^3_i,\nonumber\\
\end{eqnarray}
where $(x^3_8)_i$ are given by
\begin{eqnarray}\label{x37}
&&(x^3_8)_1=\frac{512 (21 + 11 d + d^2)}{d (1 + d)^2 (2880 - 2304 d
- 1796 d^2 + 976 d^3 + 389 d^4 - 116 d^5 -
   34 d^6 + 4 d^7 + d^8)},\nonumber\\
&&(x^3_8)_2=-\frac{1536 (21 + 11 d + d^2)}{d (1 + d)^2 (2880 - 2304
d - 1796 d^2 + 976 d^3 + 389 d^4 - 116 d^5 -
   34 d^6 + 4 d^7 + d^8)},\nonumber\\
&&(x^3_8)_3=\frac{1152 (-24 + d + 4 d^2 + d^3)}{d (1 + d)^2 (2880 -
2304 d - 1796 d^2 + 976 d^3 + 389 d^4 - 116 d^5 -
   34 d^6 + 4 d^7 + d^8)},\nonumber\\
&&(x^3_8)_4=\frac{1024 (-72 + 24 d + 23 d^2 + 4 d^3)}{d (1 + d)^2
(2880 - 2304 d - 1796 d^2 + 976 d^3 + 389 d^4 - 116 d^5 -
   34 d^6 + 4 d^7 + d^8)},\nonumber\\
&&(x^3_8)_5=-\frac{1536 (72 - 24 d - 2 d^2 + 7 d^3 + d^4)}{d (1 +
d)^2 (2880 - 2304 d - 1796 d^2 + 976 d^3 + 389 d^4 - 116 d^5 -
   34 d^6 + 4 d^7 + d^8)},\nonumber\\
&&(x^3_8)_6=-\frac{2304 (-24 + d + 4 d^2 + d^3)}{d (1 + d)^2 (2880 -
2304 d - 1796 d^2 + 976 d^3 + 389 d^4 - 116 d^5 -
   34 d^6 + 4 d^7 + d^8)},\nonumber\\
&&(x^3_8)_7=\frac{64 (144 - 264 d - 25 d^2 + 33 d^3 + 3 d^4 +
d^5)}{d (1 + d)^2 (2880 - 2304 d - 1796 d^2 + 976 d^3 + 389 d^4 -
116 d^5 -
   34 d^6 + 4 d^7 + d^8)},\nonumber\\
&&(x^3_8)_8=\frac{512 (-72 + 168 d - 25 d^2 - 21 d^3 + 3 d^4 +
d^5)}{(-3 + d) d (1 + d)^2 (5 + d) (-16 + d^2) (-4 + d^2) (-3 + 2 d
+ d^2)}.
\end{eqnarray}

\end{document}